\documentclass{PoS}

\usepackage{graphicx}
\usepackage{amsmath}
\usepackage{amssymb}

\newcommand{\be}{\begin{equation}}
\newcommand{\ee}{\end{equation}}
\newcommand{\bea}{\begin{eqnarray}}
\newcommand{\eea}{\end{eqnarray}}
\newcommand{\bi}{\begin{itemize}}
\newcommand{\ei}{\end{itemize}}
\newcommand{\ben}{\begin{enumerate}}
\newcommand{\een}{\end{enumerate}}
\newcommand{\bt}{\begin{tabbing}}
\newcommand{\et}{\end{tabbing}}
\newcommand{\nn}{\nonumber}

\newcommand{\calO}{{\mathcal O}}

\newcommand{\bfp}{{\bf p}}
\newcommand{\bfpp}{{{\bf p}^\prime}}

\newcommand{\bfx}{{\bf x}}
\newcommand{\bfxp}{{{\bf x}^\prime}}
\newcommand{\bfxpp}{{{\bf x}^{\prime\prime}}}

\newcommand{\bfz}{{\bf 0}}
\newcommand{\dt}{{\Delta x_4}}
\newcommand{\dtp}{{\Delta x_4^\prime}}
\newcommand{\fpzDP}{{f_{\{+,0\}}^{DP}}}
\newcommand{\fpDP}{{f_+^{DP}}}
\newcommand{\fmDP}{{f_-^{DP}}}
\newcommand{\fzDP}{{f_0^{DP}}}
\newcommand{\fpzDpi}{{f_{\{+,0\}}^{D\pi}}}
\newcommand{\fpDpi}{{f_+}^{D\pi}}
\newcommand{\fpDK}{{f_+}^{DK}}
\newcommand{\fpDpiK}{{f_+}^{D\pi(K)}}
\newcommand{\tsrc}{{x_{4,\rm src}}}

\title{
   \begin{picture}(0,0)(0,0)%
   \put(350,75){\makebox(0,0)[l]{\textnormal{\normalsize KEK-CP-350}}}%
   \end{picture}%
   $D$ meson semileptonic decays in lattice QCD with M\"obius domain-wall quarks
}

\ShortTitle{
   $D$ meson semileptonic decays in lattice QCD with M\"obius domain-wall quarks
}

\author{
   JLQCD Collaboration:
   \speaker{T.~Kaneko}$^{a,b}$\thanks{E-mail: takashi.kaneko@kek.jp},
   B.~Fahy$^a$, 
   H.~Fukaya$^c$, 
   S.~Hashimoto$^{a,b}$
   \\
   \\
   \\
   \llap{$^a$}
   High Energy Accelerator Research Organization (KEK),
   Ibaraki 305-0801, Japan 
   \\
   \llap{$^b$}
   School of High Energy Accelerator Science,
   SOKENDAI (The Graduate University for Advanced Studies),
   Ibaraki 305-0801, Japan
   \\
   \llap{$^c$}
   Department of Physics, Osaka University, 
   Osaka 560-0043, Japan
}

\abstract{
We report on our study of the $D$ meson semileptonic decays 
in 2+1 flavor lattice QCD. Gauge ensembles are generated
at three lattice cutoffs up to 4.5~GeV
and with pion masses as low as 300~MeV.
We employ the M\"obius domain-wall fermion action for both light and charm quarks. 
We report our preliminary results for the vector and scalar form factors
and discuss their dependence on 
the momentum transfer, quark masses and lattice spacing.
}

\FullConference{
   34th annual International Symposium on Lattice Field Theory\\
   24-30 July 2016\\
   University of Southampton, UK
}

\begin{document}


\section{Introduction}


The $D \!\to\! \pi l \nu$ and $D \!\to\! K l \nu$ semileptonic decays provide
a precise determination of the Cabibbo-Kobayashi-Maskawa matrix elements
$|V_{cd}|$ and $|V_{cs}|$, 
and play an important role in the search for new physics 
in the charm sector~\cite{NP}.
The vector and scalar form factors $\fpzDP$ 
describe non-perturbative QCD effects,
and are defined from the relevant hadronic matrix elements as 
\bea
   \langle P(p^\prime) | V_\mu | D(p) \rangle 
   & = & 
   (p+p^\prime)_\mu \fpDP(t) + (p-p^\prime)_\mu \fmDP(t),
   \label{eqn:intro:ME}
   \\
   \fzDP(t)
   & = & 
   \fpDP(t) + \frac{t}{M_D^2-M_P^2} \fmDP(t),
   \label{eqn:intro:f0}
\eea
where 
$P$ specifies the light meson ($P\!=\!\pi,K$)
and $t=(p-p^\prime)^2$ is the momentum transfer.
The current accuracy of $|V_{cd}|$ and $|V_{cs}|$ is limited 
by the theoretical uncertainty of $\fpzDP$~\cite{CKM16}.
Lattice QCD is the only known method to calculate $\fpzDP$
with controlled and systematically-improvable uncertainties.


\begin{table}[b]
\begin{center}
\caption{
   Simulation parameters.
}
\vspace{2mm}
\label{tbl:intro:status}
\begin{tabular}{l|llll|l}
   \hline 
   lattice parameters 
   & $m_{ud}$ & $m_s$ & $M_\pi$[MeV] & $M_K$[MeV] & $N_\tsrc$ 
   \\ \hline
   $\beta\!=\!4.17$, \ 
   $a^{-1}\!=\!2.453(4)$, \ 
   $32^3\!\times\!64\!\times\!12$
   & 0.0190 & 0.0400 & 499(1) & 618(1) & 2
   \\
   & 0.0120 & 0.0400 & 399(1) & 577(1) & 2
   \\
   & 0.0070 & 0.0400 & 309(1) & 547(1) & 4
   \\ \cline{2-6}
   & 0.0190 & 0.0300 & 498(1) & 563(1) & 2
   \\ \hline
   $\beta\!=\!4.35$, \ 
   $a^{-1}\!=\!3.610(9)$, \ 
   $48^3\!\times\!96\!\times\!8$
   & 0.0120 & 0.0250 & 501(2) & 620(2) & 2
   \\
   & 0.0080 & 0.0250 & 408(2) & 582(2) & 2
   \\
   & 0.0042 & 0.0250 & 300(1) & 547(2) & 4
   \\ \cline{2-6}
   & 0.0120 & 0.0180 & 499(1) & 557(2) & 2
   \\ \hline
   $\beta\!=\!4.47$, \ 
   $a^{-1}\!=\!4.496(9)$, \ 
   $64^3\!\times\!128\!\times\!8$
   & 0.0030 & 0.0150 & 284(1) & 486(1) & 1
   \\ \hline
\end{tabular}
\end{center}
\vspace{0mm}
\end{table}

In this article, 
we report on our calculation of these form factors in $N_f\!=\!2+1$ QCD
with the tree-level improved Symanzik gauge action
and the M\"obius domain wall quark action~\cite{MDWF}.
Numerical simulations are carried out at three lattice cutoffs
$a^{-1}\!\sim\!2.5$, 3.6 and 4.5 GeV.
On such fine lattices, we employ the domain-wall action
also for charm quarks.
The simulated values of $m_{ud}$, 
that is the mass of the degenerate up and down quarks,
cover a range of the pion mass 
$300~\mbox{MeV}\!\lesssim\!M_\pi\!\lesssim\!500$~MeV.
We take a strange quark mass $m_s$ close to its physical value.
An additional value of $m_s$ is simulated at certain choices of $(a,m_{ud})$
in order to study the $m_s$ dependence of the form factors.
The charm quark mass is set to its physical value 
determined from the $D$ meson spectrum~\cite{Lat15:Fahy}.
The physical charm quark mass extracted from the same set of 
simulations is $m_c(3~\mbox{GeV})\!=\!1.003(10)$\,GeV,
which is consistent with the present world average~\cite{mc:JLQCD}.
Our simulation parameters are summarized in Table~\ref{tbl:intro:status}.
After the previous report~\cite{Lat15:Suzuki},
we extend our simulation to the two finer lattices and 
improve our measurement method to reduce the statistical uncertainty.

At each simulation point, 
our lattice size satisfies a condition $M_\pi L \!\gtrsim\!4$ 
to control finite volume effects, 
and we accumulate 5,000 Molecular Dynamics time.
Chiral symmetry is preserved to good accuracy 
by choosing the sign function approximation and the kernel operator
in the 4-dimensional effective action~\cite{MDWF:JLQCD}.
The residual mass is suppressed to $O(1~\mbox{MeV})$ at the coarsest lattice,
and even smaller $\lesssim\!0.2$~MeV at finer lattices
with moderate sizes in the fifth dimension $\sim\!10$.

\section{Calculation of form factors}


We calculate the three-point function
\bea
    C^{DP}_{V_\mu}
    (\bfp,\bfp^\prime;\dt,\dtp)
    & = &
    \frac{1}{N_s^3\,N_\tsrc}
    \sum_{\tsrc}
    \sum_{\bfx,\bfxp,\bfxpp} 
    \langle 
       {\calO}_P(\bfxpp,\tsrc+\dt+\dtp) 
    \nn \\
    &&
    \hspace{10mm}
       \times
       V_\mu(\bfxp,\tsrc+\dt)
       {\calO}_D^{\dagger}(\bfx,\tsrc)
    \rangle
    e^{-i\bfpp\left( \bfxpp - \bfxp \right)}
    e^{-i\bfp\left( \bfxp - \bfx \right)},
    \hspace{10mm}
    \label{eqn:sim:msn_corr:msn_corr_3pt}  
\eea
where 
$N_s$ is the spatial lattice size and 
$\bfp^{(\prime)}$ represents the momentum of the initial (final) meson.
We apply a Gaussian smearing to the meson interpolating operators
${\calO}_{\{\pi,K,D\}}$.
The temporal coordinate of the source operator is denoted by $\tsrc$,
and $\dt^{(\prime)}$ represents 
the temporal separation between the source (sink) operator 
and the vector current $V_\mu$.


In this study,
we calculate $C^{DP}_{V_\mu}(\bfp,\bfp^\prime;\dt,\dtp)$ 
by varying $\dt$ with $\dt\!+\!\dtp$ kept fixed. 
Its physical length is the same for the three cutoffs 
and is chosen as $\dt\!+\!\dtp\!=\!28a$ at $\beta\!=\!4.17$~\cite{Lat15:Suzuki}.
The $D$ meson is at rest ($\bfp\!=\!\bfz$),
and we simulate four different values of the momentum transfer $t$
with light meson momenta $|\bfpp|^2\!=\!0,1,2,3$ in units of $(2\pi/L)^2$.
For $P\!=\!\pi(K)$,
the minimum value of the momentum transfer is typically 
$t_{\rm min}\!\approx\!0.3(0.2)\,\mbox{GeV}^2$,
while the maximum is $t_{\rm max}\!\approx\!2.6(1.8)\,\mbox{GeV}^2$.


We also calculate two-point functions of $\pi$, $K$ and $D$ mesons.
The amplitudes of the correlators are extracted
by the following exponential fits in terms of $\dt$ 
\bea
   C_{V_\mu}^{DP}(\bfp,\bfpp;\dt,\dtp)
   & = &
   A_{V_\mu}^{DP}(\bfp,\bfpp) e^{-E_D(\bfp)\dt}e^{-E_P(\bfpp)\dtp}
   \hspace{5mm}
   (P\!=\!\pi,K),
   \\ 
   C^Q(\bfp;\dt)
   & = &
   B^Q(\bfp) e^{-E_Q(\bfp)\dt}
   \hspace{5mm}
   (Q\!=\!\pi,K,D).
   \label{eqn:amp}
\eea
Here the meson energies $E_{\{\pi,K,D\}}$ are estimated
from their rest masses~\cite{Lat15:Fahy}
and the dispersion relation in the continuum limit.
The matrix elements are given as
\bea
   \langle P(\bfpp) | V_\mu | D(\bfp) \rangle
   & = &
   2 Z_V
   \sqrt{
      \frac{ E_D(\bfp)\,E_P(\bfpp)\, |A_{V_\mu}^{DP}(\bfp,\bfpp)|^2 }
           { B^D(\bfp)\,B^P(\bfpp) }
   },
\eea
where we use the renormalization factor $Z_V$
non-perturbatively calculated in Ref.~\cite{NPR:JLQCD}.
The relevant semileptonic form factors are then extracted
via Eqs.~(\ref{eqn:intro:ME}) and (\ref{eqn:intro:f0}).



In order to improve the statistical accuracy,
we average the three- and two-point functions
over the locations of the source operator.
As for the temporal location, 
we repeat our measurement over $N_\tsrc$ different values of $\tsrc$.
Our choice of $N_\tsrc$ is summarized in Table~\ref{tbl:intro:status}.
An important improvement from Ref.~\cite{Lat15:Suzuki}
is to average over the spatial coordinates $\bfx$ as well
by putting the Gaussian source operator associated with a $Z_2$ noise
at each lattice site at a given time-slice $\tsrc$.

\begin{figure}[t]
\begin{center}
   \includegraphics[angle=0,width=0.60\linewidth,clip]%
                   {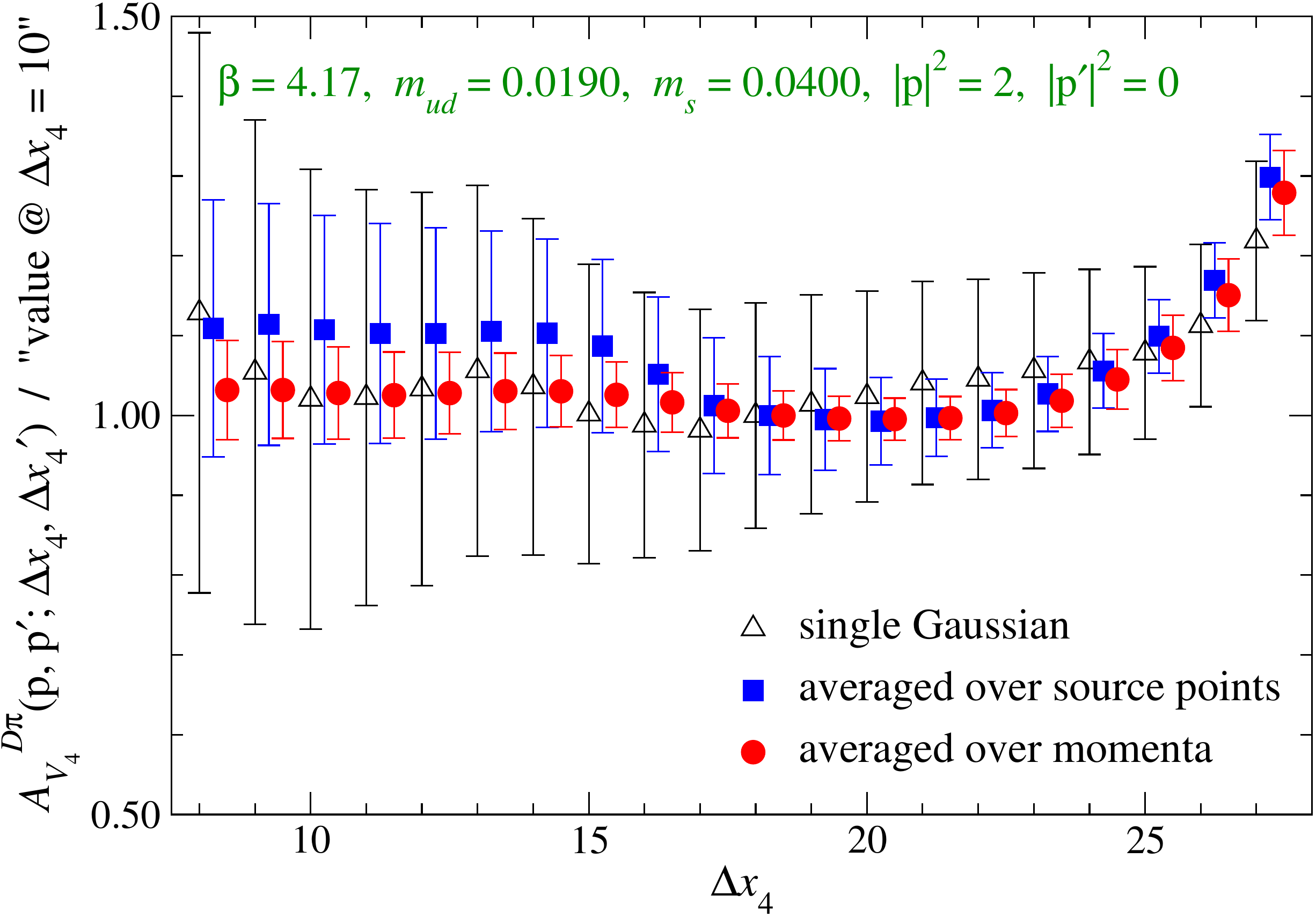}

   \vspace{0mm}
   \caption{
      Effective value of amplitude $A_{V_4}^{D \pi}(\bfp,\bfpp)$
      as a function of $\dt$.
      We plot data with $|\bfp|\!=\!2$ and $|\bfpp|^2\!=\!0$
      at $\beta\!=\!4.17$ and $(m_{ud},m_s)\!=\!(0.0190,0.0400)$.
      The open triangles show data with a ``local'' Gaussian source
      and a single choice of $\bfp$.
      The blue squares and red circles are obtained
      by averaging over the location of the source operator
      and then over 12 $\bfp$'s.
      All data are normalized by their value at $\dt\!=\!18$.
   }
   \label{fig:ff:comp}
\end{center}
\vspace{0mm}
\end{figure}


Figure~\ref{fig:ff:comp} compares our results for the amplitude $A_{V_4}^{D\pi}$
with different measurement setups
on our coarsest lattice at the heaviest sea quark masses.
We observe about 30\,\% reduction of the statistical error
by averaging over 2 $t_{\rm src}$'s
and an additional 30\,\% reduction by averaging over $\bfx$:
about a factor of two improvement in total.
Averaging over $\bfp$ further improves the statistical accuracy:
at $|\bfp|^2\!=\!2$, for instance,
about factor of two improvement by averaging over 12 $\bfp$'s.
The typical statistical accuracy is
1\,--\,2\,\% at $t_{\rm max}$ and $M_\pi\!\sim\!500$~MeV,
and 6\,--\,9\,\% at $t_{\rm min}$ and $M_\pi\!\sim\!300$~MeV.


\section{Momentum transfer dependence}


We parametrize the momentum transfer dependence
of the form factors in terms of the so-called $z$ parameter~\cite{z}
\bea
   z(t,t_0)
   & = &
   \frac{ \sqrt{t_+-t} - \sqrt{t_+-t_0} }
        { \sqrt{t_+-t} + \sqrt{t_+-t_0} },
   \label{eqn:q2-dep:z}
\eea
where $t_+\!=\!(M_D+M_P)^2$ represents the $DP$ threshold ($P\!=\!\pi$, $K$).
The free parameter $t_0$ is chosen so that
our simulated region $t\!\in\![t_{\rm min},t_{\rm max}]$
is mapped into a shortest segment 
$z\!\in\!\left[-|z|_{\rm max},+|z|_{\rm max}\right]$
centered at the origin.
Typical size of the $z$ parameter is $|z|_{\rm max}\!\lesssim\!0.2$.

\begin{figure}[t]
\begin{center}
   \includegraphics[angle=0,width=0.60\linewidth,clip]%
                   {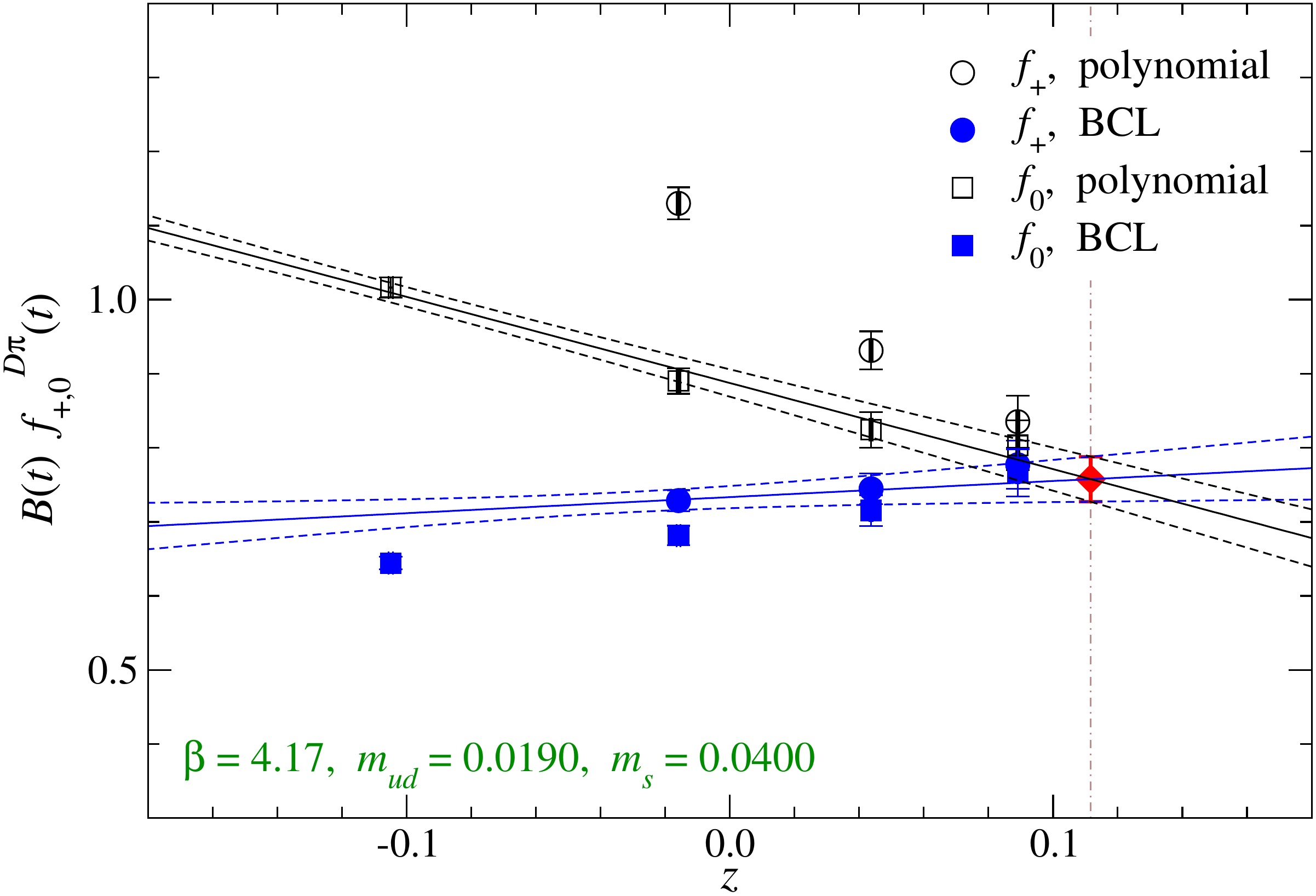}

   \vspace{0mm}
   \caption{
      Plot of $B(t)\,f_{\{+,0\}}^{D\pi}(t)$ as a function of $z$
      at $\beta\!=\!4.17$ and $(m_{ud},m_s)\!=\!(0.0190,0.0400)$.
      Circles and squares are data for $f_+$ and $f_0$, respectively.
      Filled symbols show
      $(1-t/M_{D^*_{(0)}}^2) f_{+(0)}^{D\pi}(t)$
      for the BCL parametrization, 
      whereas 
      open symbols for the polynomial expansion 
      are just the form factors themselves.
      A simultaneous fit to $f_+$ (filled symbols)
      and $f_0$ (open symbols) is shown by solid and dashed lines.
      The vertical dot-dashed line represents $z$ corresponding $t\!=\!0$,
      and the diamond is the value extrapolated to $t\!=\!0$.
   }
   \label{fig:q2-dep:ff_vs_z}
\end{center}
\vspace{0mm}
\end{figure}


The momentum transfer dependence of the form factors are then
parametrized by using this small parameter as 
\bea
   \fpzDP(t)
   & = &
   \frac{1}{B(t)} \sum_{k=0}^{N_{\{+,0\}}} a_{\{+,0\},k}\, z^k.
   \label{eqn:q2-dep:ff_vs_z}
\eea
In this preliminary analysis,
we test two choices of the factor $B(t)$. 
In the so-called Bourrely-Caprini-Lellouch (BCL) parametrization~\cite{BCL}
with 
\bea
   B(t) = 1 - \frac{t}{M_{\rm pole}^2},
   \label{eqn:q2-dep:ff_vs_z:bcl}
\eea   
possibly small deviation from the lowest pole contribution $1/B(t)$
is expanded in terms of $z$.
We also test a naive polynomial expansion of the form factors themselves
with 
\bea
   B(t) = 1.
   \label{eqn:q2-dep:ff_vs_z:poly}
\eea


Figure~\ref{fig:q2-dep:ff_vs_z} shows 
$z$-dependence of a quantity $B(t)\fpzDpi(t)$ to be expanded in terms of $z$.
Namely, $B(t)f_{+(0)}^{D\pi}(t)\!=\!(1-t/M_{D^*_{(0)}}^2) f_{+(0)}^{D\pi}(t)$ 
for the BCL parametrization with Eq.~(\ref{eqn:q2-dep:ff_vs_z:bcl}),
whereas it is just the form factor for the polynomial expansion 
with Eq.~(\ref{eqn:q2-dep:ff_vs_z:poly}).


For $\fpDP$, 
we use the vector meson masses $M_{D_{(s)}^*}$ calculated at the simulation points,
which are well below the threshold $t_+$. 
We observe that the $z$ dependence of $B(t)\fpDP(t)$ is significantly reduced
by switching from the polynomial expansion (\ref{eqn:q2-dep:ff_vs_z:poly})
to the BCL parametrization (\ref{eqn:q2-dep:ff_vs_z:bcl}).
This suggests that the vector meson dominance (VMD) hypothesis is
a reasonably good approximation of $\fpDP$,
and we can expand the small deviation from the VMD
in terms of small $z$.
In this study, we test
two BCL parametrization
including the linear ($N_+\!=\!1$) and quadratic terms ($N_+\!=\!2$).


We have not yet calculated the scalar meson masses $M_{D_{(s)0}^*}$,
and hence it is not clear whether there exist
corresponding isolated poles below $t_+$ at simulated $M_\pi$'s.
In this analysis,
we employ the simple linear expansion (\ref{eqn:q2-dep:ff_vs_z:poly})
for $\fzDP$.
We also test 
the BCL parametrization ($N_0\!=\!1$) 
with the experimental value of $M_{D_{(s)0}^*}$
by assuming its mild dependence on $m_{ud}$.


We estimate the normalization $\fpDP(0)\!=\!\fzDP(0)$
from the simultaneous fit
using the BCL parametrization with $N_+\!=\!1$ for $\fpDP$
and the linear parametrization ($N_0\!=\!1$) for $\fzDP$.
We also test above-mentioned alternative forms
to estimate the systematic uncertainty of the extrapolation to $t\!=\!0$.
As shown in Fig.~\ref{fig:q2-dep:ff_vs_z}, however, 
the $z$ dependence of our results is mild 
except the polynomial parametrization for $\fpDP$. 
The systematic uncertainty is not large compared to
the statistical accuracy.


\section{Continuum and chiral extrapolation}

\begin{figure}[t]
\begin{center}
   \includegraphics[angle=0,width=0.48\linewidth,clip]%
                   {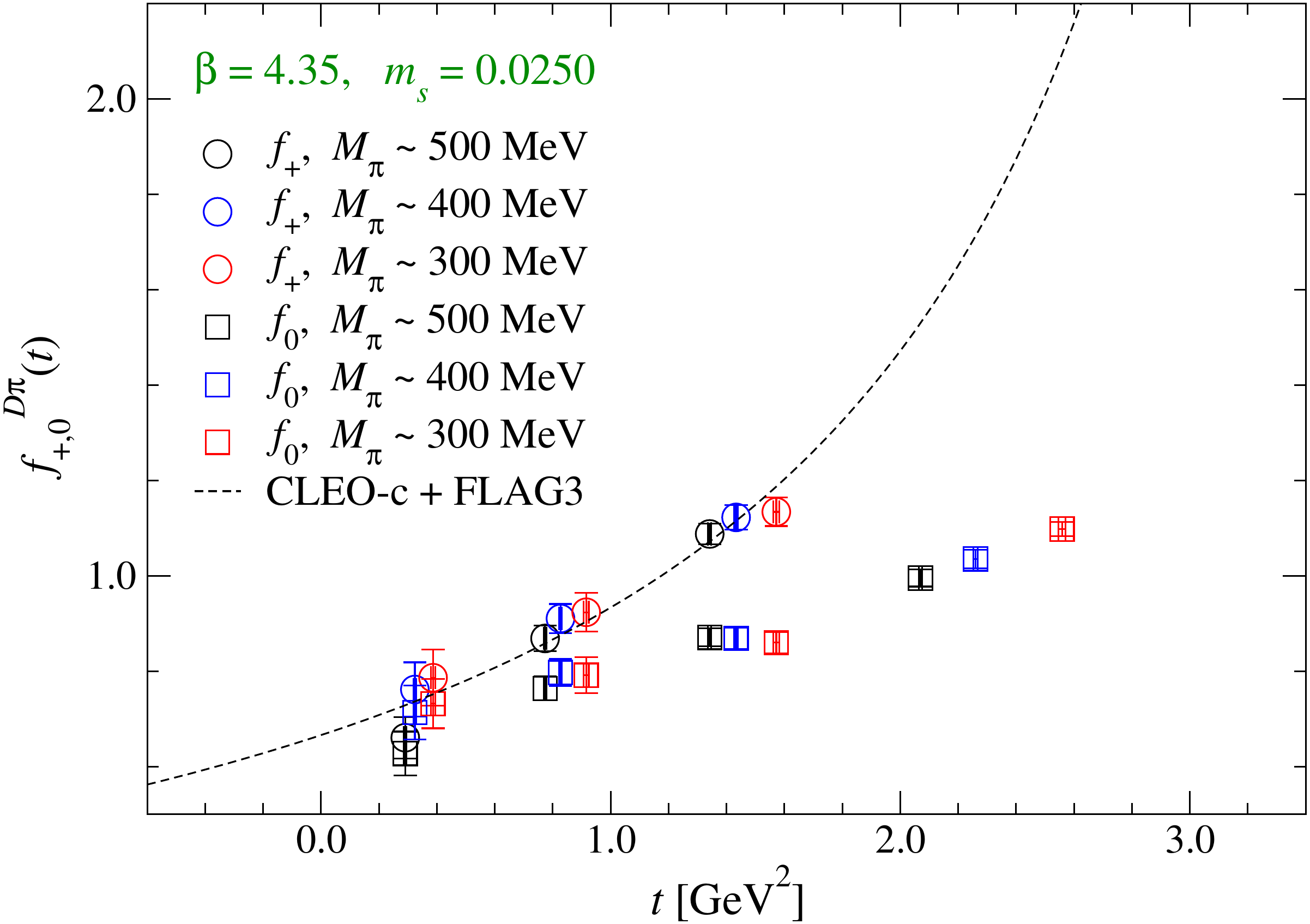}
   \hspace{3mm}
   \includegraphics[angle=0,width=0.48\linewidth,clip]%
                   {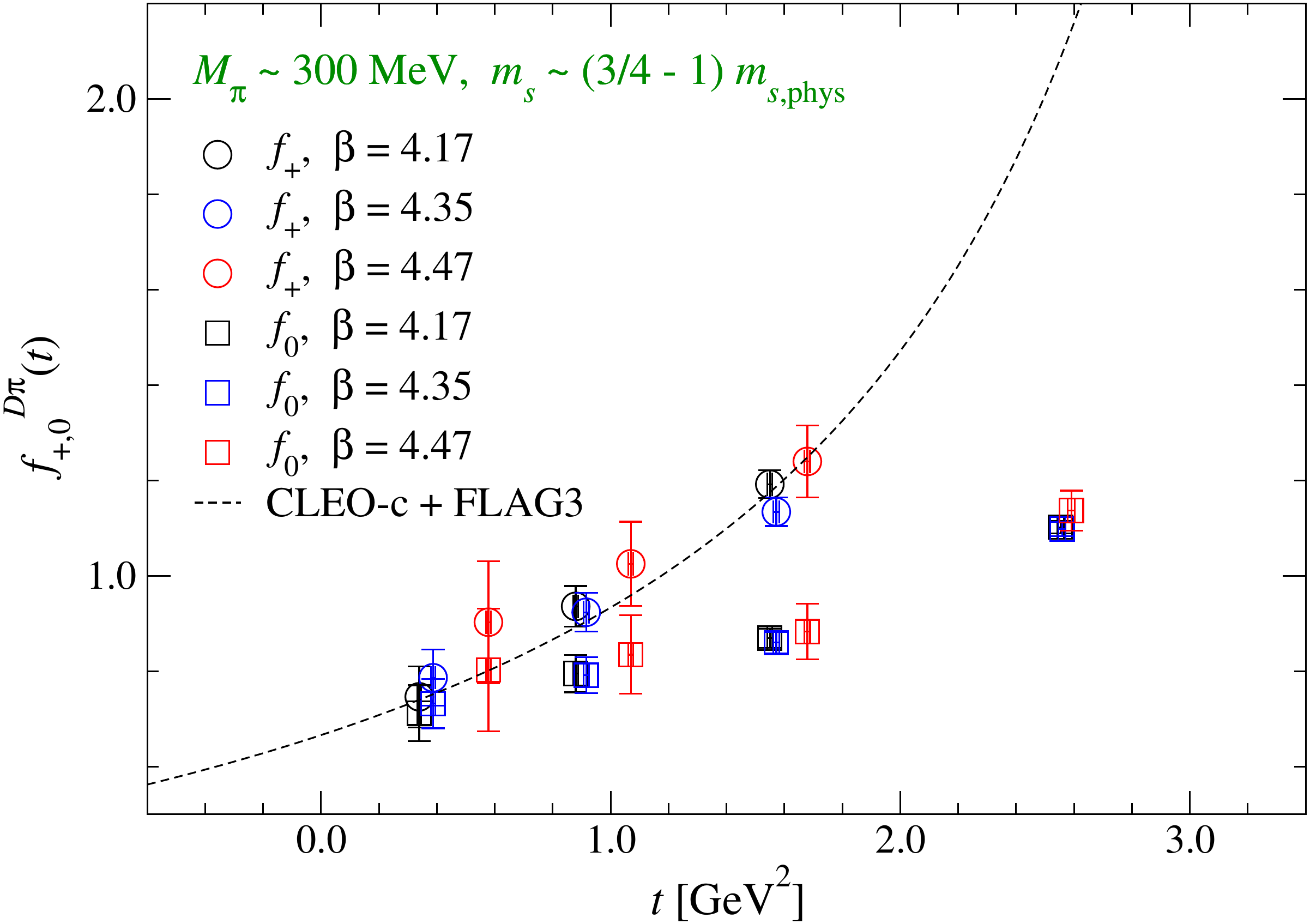}
   \vspace{0mm}
   \caption{
      Comparison of $\fpzDpi$
      among different $M_\pi$'s (left panel) and 
      different $a$'s (right panel). 
      We plot data at $\beta\!=\!4.35$ and $m_s\!=\!0.0250$ in the left panel,
      whereas the right panel shows data at $M_\pi\!\sim\!300$~MeV
      and larger $m_s$.
      We also plot
      the Becirevic-Kaidalov parametrization~\cite{BK}
      of the CLEO-c data of $\fpDpi(t)$~\cite{CLEO-c}
      combined with an average of recent lattice estimates 
      of $\fpDpi(0)$~\cite{FLAG3}.
   }
   \label{fig:cont+chiral_fit:mud-dep_a-dep}
\end{center}
\vspace{0mm}
\end{figure}


In Fig.~\ref{fig:cont+chiral_fit:mud-dep_a-dep}, 
we compare $\fpzDpi$ at different pion masses (left panel)
and at different lattice spacings (right panel).
The reasonable consistency in both panels
suggests a mild dependence on $M_\pi$ and $a$. 
We note that the decay constants $f_{D_{(s)}}$ also have 
small discretization errors 
with our choice of the lattice action and cutoffs~\cite{Lat16:Fahy}.

\begin{figure}[t]
\begin{center}
   \includegraphics[angle=0,width=0.48\linewidth,clip]%
                   {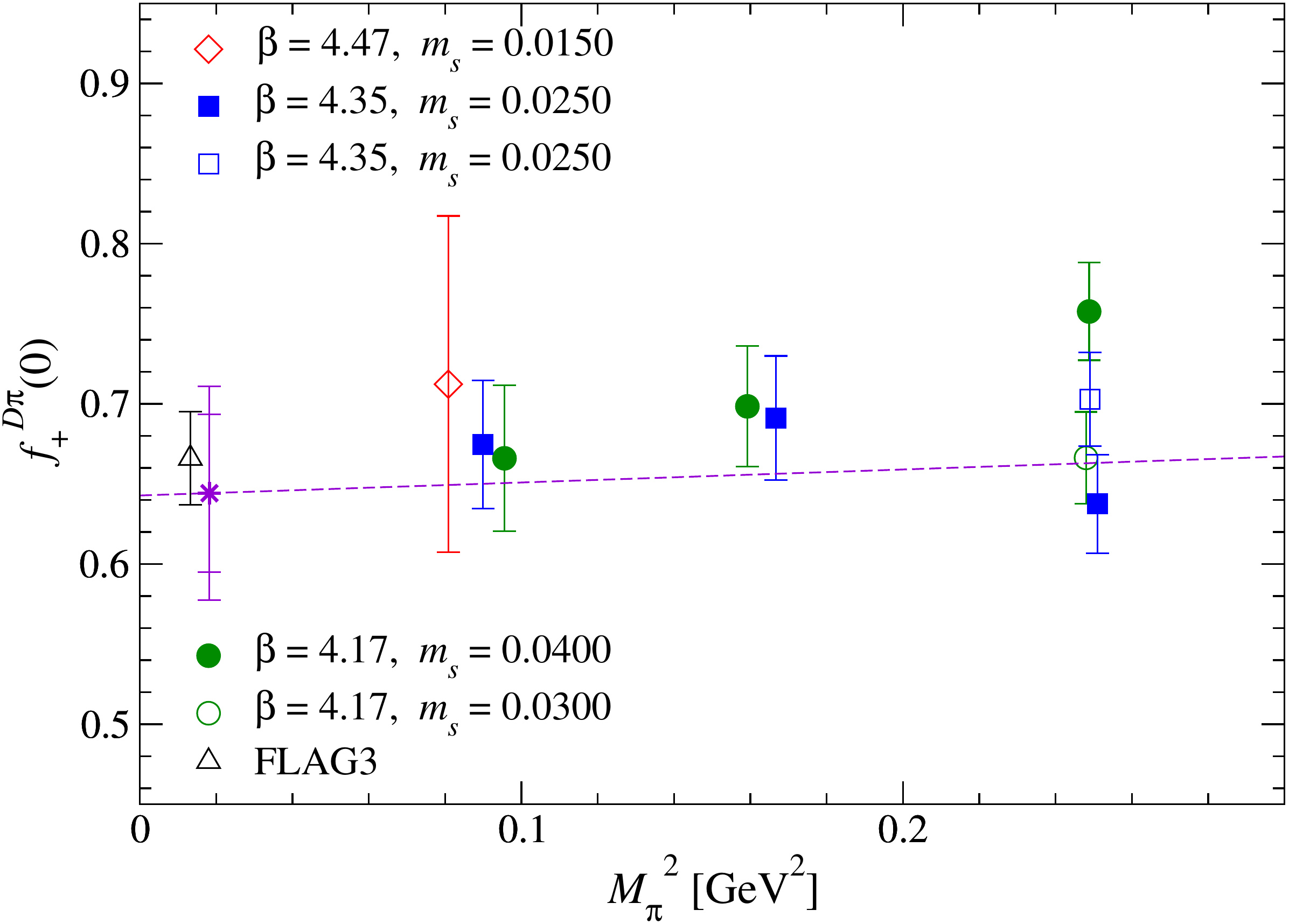}
   \hspace{3mm}
   \includegraphics[angle=0,width=0.48\linewidth,clip]%
                   {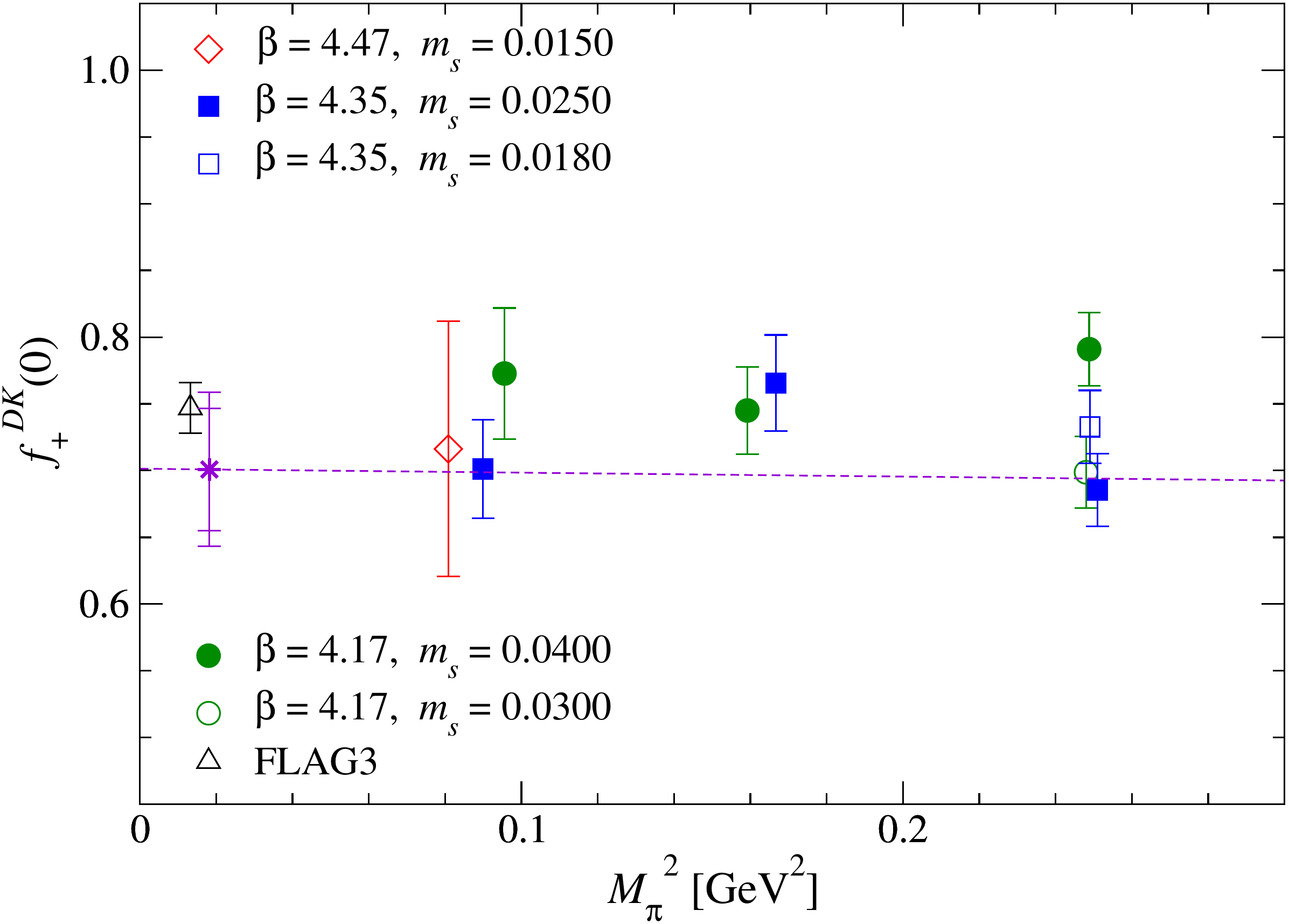}
   \vspace{0mm}
   \caption{
      Continuum and chiral extrapolation of 
      $\fpDpi(0)$ (left panel) and $\fpDK(0)$ (right panel).
      Data at different $a$'s and $m_s$'s are 
      plotted by different symbols as a function of $M_\pi^2$.
      The dashed lines show the fit line in the continuum limit
      and at the physical strange quark mass.
      The value extrapolated to the physical point is plotted by the stars.
      We also plot averages of recent lattice estimates~\cite{FLAG3}
      by the triangles.
   }
   \label{fig:cont+chiral_fit:lin}
\end{center}
\vspace{0mm}
\end{figure}

In this preliminary analysis, therefore,
we parametrize the $a$, $m_{ud}$ and $m_s$ dependences of $\fpDP(0)$
by the following simple linear form
\bea
   \fpDP(0)
   & = & 
   c^{DP}
 + c_{a}^{DP} a^2
 + c_{\pi}^{DP} M_\pi^2
 + c_{\eta_s}^{DP} M_{\eta_s}^2,
\eea
where $M_{\eta_s}^2\!=\!2M_K^2-M_\pi^2$.
This continuum and chiral extrapolation is plotted 
in Fig~\ref{fig:cont+chiral_fit:lin}.
We obtain $\chi^2/{\rm d.o.f}\!\sim\!1.6$\,--\,1.8.
All the coefficients $c_{\{a,\pi,\eta_s\}}^{DP}$ 
have $\gtrsim\,75$\,\% statistical error:
namely, consistent with zero
as expected from the good consistency 
in Fig.~\ref{fig:cont+chiral_fit:mud-dep_a-dep}.
This fit is therefore not sensitive to higher order corrections,
and we estimate the systematic uncertainty 
from three fits 
in which one of the three coefficients $c_{\{a,\pi,\eta_s\}}^{DP}$ is set to zero.
Our preliminary estimates 
\bea
   f_+^{D\pi}(0)
   = 
   0.644(49)(27),
   \hspace{5mm}
   f_+^{DK}(0)
   =
   0.701(46)(33).
\eea
are consistent with recent lattice averages 
$f_+^{D\pi}(0)\!=\!0.666(29)$ and $f_+^{DK}(0)\!=\!0.747(19)$~\cite{FLAG3}.

\section{Summary}

In this article, we report on our lattice calculation
of the $D\!\to\!\pi$ and $D\!\to\!K$ semileptonic form factors.
We employ the M\"obius domain-wall quark action 
both for light and charm quarks,
and simulate lattice cutoffs up to 4.5~GeV.

Our preliminary results for $\fpDpiK(0)$ have uncertainty of 8 (9)\,\%. 
We expect significant improvement in the near future 
by increasing statistics ($N_\tsrc$) on the finest lattice 
and extending our measurements to smaller $M_\pi\!\sim\!230$~MeV.

We observe small discretization errors of the $D$ meson form factors
with our simulation setup. It is therefore interesting to extend 
our study to the $B$ meson semileptonic decays, 
which are being precisely measured at SuperKEKB/Belle II and LHCb
experiments.

\vspace{3mm}

Numerical simulations are performed on Hitachi SR16000 and 
IBM System Blue Gene Solution at KEK
under a support of its Large Scale Simulation Program (No.~16/17-14).
This research is supported in part by the Grant-in-Aid of the MEXT
(No.~26247043, 26400259)
and by MEXT as ``Priority Issue on Post-K computer''
(Elucidation of the Fundamental Laws and Evolution of the Universe) and JICFuS.


\end{document}